\documentclass{aa}
\usepackage{polski}
\usepackage[english]{babel}
\usepackage{graphicx}
\usepackage[usenames]{color}
\newcommand{\bhl}{}
\newcommand{\ehl}{}
\newcommand{\qp}{$Q^\prime_*$ }
\newcommand{\qpe}{Q^\prime_*}

\begin{document}

\title{A binary merger origin for inflated hot Jupiter planets}
\author{E.\ L.\ Martin$^1$, H.\ C.\ Spruit$^2$, and R.\ Tata$^3$}
\institute{Centro de Astrobiologia CSIC - INTA, Carretera Torrej\'on-Ajalvir km 4, 28850 Madrid, Spain, \and Max-Planck-Institut f\"ur Astrophysik, 85748 Garching, Germany 
\and  Instituto de Astrofisica de Canarias, calle Via Lactea, 38200 La Laguna, Tenerife, Spain}
\titlerunning{hot Jupiters from binary mergers}
\authorrunning{Martin et al.}
 
\abstract{

 We hypothesize that hot Jupiters with inflated sizes represent a separate planet formation channel,   the merging of two low-mass stars. 
We show that the abundance and properties of W~UMa stars and low mass detached binaries are consistent with their being possible progenitors. The degree of inflation of the transiting hot Jupiters correlates with their expected spiral-in life time by tidal dissipation, and this could indicate youth if the stellar dissipation parameter \qp is sufficiently low.  Several Jupiter-mass planets can form in the massive compact disk formed in a merger event. Gravitational scattering between them can explain the high incidence of excentric, inclined, and retrograde orbits. 

If the population of inflated planets is indeed formed by a merger process, their frequency should be much higher around blue stragglers than around T Tauri stars.
 \keywords{planets and satellites: formation, dynamical evolution and stability; blue stragglers; binaries: eclipsing; methods: statistical; individual: \object{CoRoT-12b}, \object{CoRoT-15b}, \object{V838 Mon}, \object{V1309 Sco}.}
}
\maketitle

\section{Introduction}

Ever since the first discoveries of Jupiter-mass planets around solar-type stars  (Mayor \& Queloz 1995; Butler \& Marcy 1996), the origins of hot Jupiters have inspired a variety of ideas.  
 Formation scenarios invoked in the literature include orbital migration, radiative stripping, planet scattering ,and secular chaos (Ford \& Rasio 2008; Ida \& Lin 2010; Naoz et al. 2011). 
Each of these mechanisms can explain some of observational properties of hot Jupiters, but not all of them. The main properties are summarized as follows: 
a) a pile up of orbital periods around three days; b) a wide range of obliquity; c) a lack of close  companion stars,  d) a mass function different than for other types of exoplanets,  e) a limited mass range of host star masses, f) a steep rise in  frequency for super-solar metallicities,  g) a frequency insensitive to stellar metallicity from solar down to very metal poor stars, and h) a wide range of planetary sizes for a given planetary mass. Together, all these observational patterns constitute  a substantial set of constrains that no single theory has been able to explain,  though planet population synthesis  (Mordasini et al. 2009) has met with success in accounting for  several of the observed features.  It is possible that  different pathways exist  to the formation of hot Jupiters. Hopefully they would leave different signatures on the planet properties that could be used to identify their origin. 
In this paper we hypothesize  a new scenario, where they form in excretion disks produced by low-mass binary mergers, which, we argue, could be the cause of the large sizes of some transiting hot Jupiters.

The radius of brown dwarfs and giant planets is determined by the interior temperatures which, for single objects, depend on the total mass,  age,  and the opacity.  At a given planetary mass and composition, all planets are expected to converge to a constant radius ($R_0$) with time, which is determined by electron degeneracy pressure, Coulomb pressure and exchange pressure in the interior (Zapolsky \& Salpeter 1969; Stevenson 1991). Since the stellar hosts of exoplanets are usually on the main sequence, it is difficult to estimate their ages, but typically they are thought to be several Gyr old, and if their planets have the same age, they should have had enough time to evolve to their degeneracy dominated configuration where $R\rightarrow R_0$. The advent of wide area transit surveys from the ground (such as WASP) and in space (CoRoT and Kepler) has brought about the discovery of over a hundred transiting exoplanets. The combination of detailed photometric transit observations and high-precision radial velocity measurements have provided mass and radius determinations for those planets. With all this wealth of data, it has become clear that our understanding of the mass-radius relationship in exoplanets is far from satisfactory.  In particular two properties of transiting exoplanets are specially troubling: (1) The large spread of planetary radii for masses around 1 Jupiter mass (from 10\% to as much a factor 2); and (2) the large size of some exoplanets, which has been termed the `radius anomaly' of hot Jupiters  (Bodenheimer et al.\ 2001; Guillot et al.\ 2006). 

Several mechanisms have been invoked to account for the radius anomaly, such as diffusion of stellar irradiation into the planet interior, enhanced opacities, inefficient convection, tidal heating (Leconte et al.\ 2009), Ohmic heating (Batygin \& Stevenson 2010), burial of heat by turbulence (Youdin \& Mitchell 2010), and strong winds (Chabrier, Leconte \& Baraffe 2010). The wide spread in radius excess (more than a factor of 2 for planets with masses around that of Jupiter) makes it unlikely that the cause of the radius anomaly lies in systematic errors in the physics of interior structure (EOS, opacities) or other factors that are just functions of  intrinsic properties of the planet. 

Recent Spitzer observations of the transiting planet WASP-12b have indicated that the atmospheric composition may be highly non-solar (Madhusudhan et al. 2011). 
Then it is possible that the scatter in the mass-radius relationship is at least partly due to variations in the atmospheric composition of hot Jupiters and in their cloud thickness (Burrows, Heng \& Nampaisarn 2011).  
However, to produce hot Jupiters as inflated as observed with enhanced atmospheric opacities a factor 10 above the solar mixture is needed for cooling and contraction to be slowed down significantly (Guillot 2008). 
The conventional planet formation models (core accretion) have indeed predicted that Jupiter-mass planets could have different chemical compositions from those of the nebulosity from which their parent star formed (Pollack \& Bodenheimer 1989). However, in these models, the higher mass planets that form in the protoplanetary disk have abundances closer to that of the central star. 
Thus, it is unclear how planets larger than Jupiter and even brown dwarfs may form with chemical abundances radically different from those of the host stars. Furthermore, the degree of inflation of hot Jupiters seems to be quite insensitive to the metallicity of the host star (Laughlin, Crismani \& Adams 2011).

The transiting brown dwarf CoRoT-15b has shown that the radius anomaly extends into the brown dwarf mass domain, making it more unlikely that the correct explanation is stellar irradiation (Bouchy et al. 2011). 
The age of the stellar host has been estimated to lie in the range 1.14 -- 3.35 Ga using evolutionary models, while the cooling age of the brown dwarf is around 0.5 Ga, although high-metallicity models could recover the coevality of the system 
(Burrows et al.\ 2011). This example illustrates that the radius anomaly could be interpreted either as an age discrepancy between the stars and their close substellar-mass companion, or 
as a difference in chemical composition.

\section{Hot Jupiter  formation in binary mergers}

A close binary with a system mass in the range of planet hosts consists of stars with convective envelopes. The magnetic activity of these stars causes them to lose angular momentum by `magnetic braking'. Tidal coupling between the stars transmits (part of) this loss to the orbit, the binary orbit becomes narrower, and the angular momentum loss speeds up. Eventually, the stars go through a merger process.  If in this process a substantial amount of mass remains in orbit around the primary, it would form  a disk in which planets could form. Such a scenario  for `second generation' planets around single main sequence stars has already been discussed by Tutukov et al. (1991, 2004), but not in the context of hot Jupiters.  The low angular momentum of the post-merger disk compared with a normal protoplanetary disk means that the orbits of these planets would naturally be close. Tidal interaction with the star would eventually cause the innermost planets to be accreted.  
Other examples of `late planet formation' scenarios in circumbinary excretion  disks are those proposed to explain the existence of planets around pulsars and highly evolved stars (Banit et al. 1993; Wolszczan \& Kuchner 2010; Perets 2011). 

Direct observational evidence that contact binaries merge to become single stars has been elusive, but one case is now known. OGLE photometric monitoring  has shown that the progenitor of 
Nova Sco 2008 (V1309 Sco) was not the expected cataclysmic variable but a contact binary with an orbital period of 1.4~d (Tylenda et al. 2010). 
Over the years leading up to the outburst, the period was observed to decrease ever more rapidly with time, as expected in a merger event. 
The nova-like behavior of the event shows that energetic mass ejection can accompany a W~UMa merger 
(at least, at the high end of the W~UMa mass distribution represented by V1309 Sco). An example of circumstellar dust formation around a stellar merger event 
could be the mid-infrared flux variability observed in V838 Mon (Wisniewski et al. 2008), although the nature of this object prior to its 2002 outburst is still a 
matter of debate.   

If a hot Jupiter planet is formed in the excretion disk produced by a binary merger, the cooling age of the planet would reflect the time elapsed since the merging event. 
 Calculations of the  life times of a contact binaries before their final merger indicate that they can be rather long (6-10 Gyr, St\k{e}pie\'n 2011). Thus, hot Jupiters produced in mergers would look much younger than their host stars.  Their host population is a few Gy old, as judged from its kinematics and their  inferred evolutionary status in individual cases (such as Lee et al. 2011). 

An obvious indicator of a  young cooling age for any planet is that it has a radius larger than an older planet with the same mass.   
Recognizably inflated hot Jupiters (e.g., 20\% above the nominal mass-radius relation)  comprise about 1/one-third of the known sample, and would need to 
be about $10^8$ yrs or younger to explain their sizes with standard evolutionary models (Baraffe et al. 2003). To examine the likelihood of this scenario we proceed \bhl by checking\ehl whether there is 
a suitable population of progenitor binaries, \bhl and by looking\ehl for a connection between  planet inflation and tidal evolution.

\subsection{Progenitor population}

If the progenitors are recognizable as such for a fraction $f$ of the age  of the host stars (3 Gy, say), the progenitors must thus be some $30f$ times more abundant than the inflated planet population and should be easily identifiable. The frequency of transiting planets found by CoRoT around G-type main sequence stars is around 1 per 2000 stars, so the recognizably inflated fraction of 1/3 has an abundance of about $1.6\,10^{-4}$ per G star. Their progenitors must then have an abundance of $\sim 5\,10^{-3}f$. An obvious candidate population is contact binaries (W~UMa stars). 

The abundance of stars classified as W~UMa on the basis of their (EW type) light curves is reasonably well established. From a complete sample of bright stars (Rucinski 2002) and  ASAS light curves (Rucinski 2006), an abundance of $2\, 10^{-3}$ per main sequence star of the same spectral type has been derived. Since they are eclipsing binaries, the selection effects of orbital inclination are probably similar to those of the inflated transiting planets. In the presently most plausible interpretation of the evolution of W~UMas (St\k{e}pie\'n  2006, 2009), they are long-lived (a few Gy), as for the estimated age of the planet host population, i.e.\  $f\approx 1$. On the basis of these numbers, contact binaries thus appear to be a possible progenitor population, although they may not be the only relevant possibility as discussed below. 

The outcome of close binary evolution (e.g.\ Ritter 1996) is different depending on the mass ratio, total mass, and the evolutionary status of the components. The primary star can be brought (or kept) in synchronous rotation with the orbit if the mass ratio $q=M_2/M_1$ of the secondary is large enough, and the stars close enough for tidal interaction to be effective. Ongoing angular momentum loss from the system by magnetic braking narrows the orbit until one of the stars fills its Roche lobe and mass transfer starts. If, on the other hand, the orbital angular momentum of the secondary is too small to spin the primary up to corotation with the orbit, the orbit will shrink by tidal interaction (even if magnetic braking were ineffective) until one of the stars fills its Roche lobe. If the mass ratio is below Darwin's stability limit (`tidal instability', G.\ Darwin 1879, Hut 1980), this will happen even if the stars initially rotate in synchrony with the orbit. For low mass main sequence stars, this limit is on the order of $q=0.08$ (Rasio 1995).  

After the binary orbit has shrunk until one of the stars fills its Roche lobe, the further evolution depends on the nature of the lobe-filling star, since its internal structure determines how its size responds to loss of mass from its surface. For main sequence stars the star that first fills its lobe is the more massive one. The simplest case is when it is a fully convective, low-mass star. Its adiabatic mass-radius exponent is then negative; i.e., sudden mass loss causes it to expand. Mass loss then causes it to overfill its Roche lobe, and the mass loss rate increases exponentially, with the final stages happening on a dynamical (orbital) time scale (cf.\ Ritter 1988). Since the receiving star does not fill its Roche lobe yet and, being of lower mass, is also fully convective with a negative mass-radius exponent, it can receive a large amount of mass from the primary before also filling its Roche lobe. The SPH simulations by Rasio and Shapiro (1995) and grid based simulations (D'Souza et al.\ 2006) show that the final merger of such binaries happens on a time scale of some ten orbits. 

The situation is more complicated when the mass-radius exponent of the primary is positive or becomes positive after an initial phase of dynamical mass transfer (St\k{e}pie\'n  2006). A longer  phase of mass transfer on the thermal time scale then takes place, until the mass ratio of the system has reversed, as in Algol type binaries. For primary masses on the order of $1\ M_\odot$, the star is likely to have evolved a (small) helium core during the magnetic braking period that brought the binary into contact. St\k{e}pie\'n (2006) shows that this allows binaries in the mass range of W~UMa stars to settle into a stable contact configuration, which lasts for a few Gy until angular momentum loss by magnetic braking causes the stars to finally merge. This solves a longstanding puzzle in the theory of contact binaries and explains their high abundance. 

\begin{figure}[t]
\hfil\includegraphics[width=0.8 \linewidth]{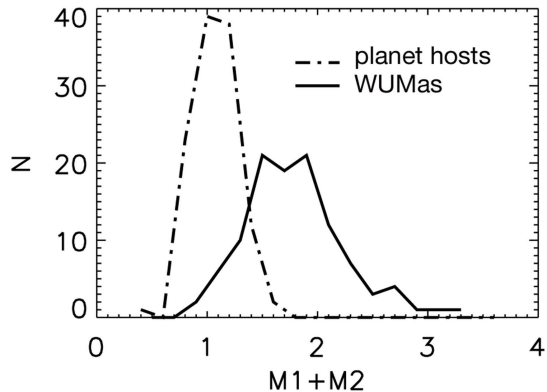}\hfil
\caption{Masses of the host stars of 103 transiting planets  for which mass and radius of star and planet are given in  the `Extrasolar Planets Encyclo\-paedia' compared with the masses of W~UMa binaries  from Gazeas \& St\k{e}pie\'n (2008).}
\label{massD}
\end{figure}

The nature and abundance of W~UMas makes them a plausible candidate population for making hot Jupiters, but they are probably not the only ones. The distribution of (total) mass of W~UMas  in the catalog of Gazeas \& St\k{e}pie\'n (2008) peaks at $1.5 - 2 M_\odot $, the distribution of known host masses of transit planets around $1.1 M_\odot $. 
In Fig.\ \ref{massD} we compare the distribution of total masses of W~UMa stars with that of transiting hot Jupiter host stars.  The overlap between the distributions is in fact only modest. The V1309 Sco event shows that some mass can be lost in the final merger, although loss as large as several tenths of a solar mass seems unlikely. It is thus worth exploring other channels. We propose here that the additional channel is in fact the direct merger of low-mass binaries ($M1+M2 \la 1.2 M_\odot$) on a dynamical time scale as discussed above. There is no obvious reason why such systems, at masses of $0.8 - 1.2 M_\odot $, would be formed at rates much less than the  $1- 2.5 M_\odot $ binaries that end up becoming W~UMas. Their lower luminosity, and the absence of the extended contact phase that makes W~UMas stand out, would make them a much less prominent population.  Detached binary systems that could be such progenitors are known (e.g.\ Coughlin et al.\  2011). 

\subsection{Spiral-in time scales}

\begin{figure}[t]
\hfil\includegraphics[width=1.0 \linewidth]{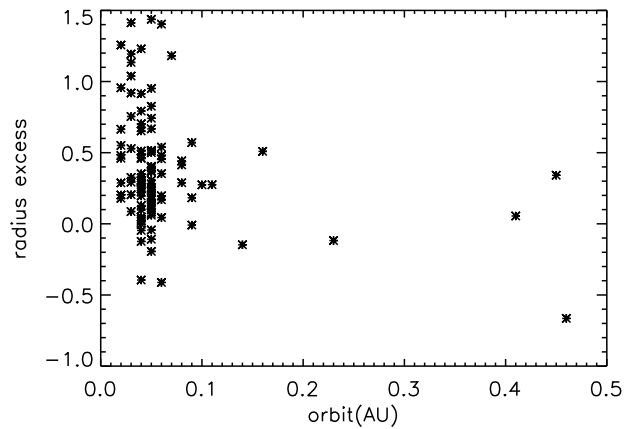}\hfil
\caption{Radius excess $\delta$ vs.\ orbital distance from the host star of 103 planets  for which mass and radius of star and planet are given in  the `Extrasolar Planets Encyclo\-paedia'.}
\label{orb}
\end{figure}

Inflated hot Jupiters orbit at distances close to their host star, typically within 0.05 AU (Fig.\ \ref{orb}).  At such distances, friction in the tides they raise on the star cause their orbits to circularize. The functional dependence of this process on system parameters is well established (Goldreich 1963; Jackson et al. 2009). The associated time scale is proportional to a  stellar tidal dissipation parameter \qp whose value is less well known since it depends on details of the interaction of the tides with the flows in the convective envelope of the host  star. By fitting the observed distribution of excentricities of planet orbits, Jackson, Greenberg \& Barnes (2008) deduce a value of $\qpe = 3\,10^5$ for this dimensionless parameter  in the host stars and $Q^\prime_{\rm p} = 3\,10^6$ in the planets. 

A much higher value of stellar dissipation parameter is predicted, however, by a quantitative theory for the dissipation of internal waves in the radiative interior of the solar type stars by Barker \& Ogilvie (2010), although it does not include the dissipation due to interaction of the tides with flows in the convective envelope, which is a more complicated process.   On the other hand, a lower  value,  $Q^\prime = 3.6\,10^4$, has been inferred for Jupiter from astrometric analysis of the 
orbital motions of the Galilean moons by  Lainey et al. (2009), and a value $Q^\prime_{\rm p} = 10^5$ was adopted for exoplanets by Weidner \& Horne (2010), who noted that the constraints of detections of moons around hot Jupiters could be used to infer a lower limit  on the planetary dissipation parameter.

\begin{figure}[t]
\hfil\includegraphics[width=0.8 \linewidth]{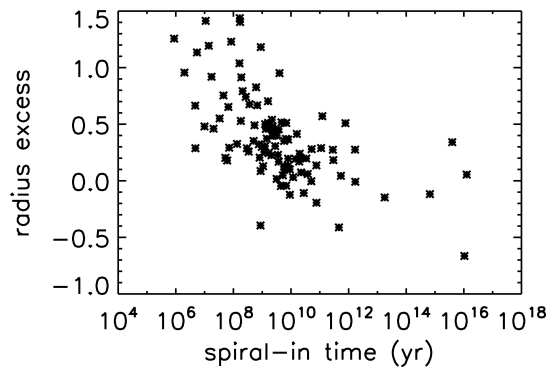}\hfil
\caption{ Radius excess $\delta$ vs.\ spiral-in time scale for the sample of Fig.\ \ref{orb} for  an assumed dissipation parameter $\qpe = 10^6$.}
\label{age1}
\end{figure}

The tides raised on the star are a significant sink of orbital angular momentum. If the mass of the companion is low, its angular momentum can be too small to keep the host star corotating with the orbit (Darwin instability), and the planet will eventually spiral into the star. The time scale for this is also known (cf.\ Debes and Jackson 2010), subject again to the uncertainty  of the  parameter $\qpe$. If mass and radius of both the planet and the host star are known, the expected life time of the planet before spiral-in can be calculated.  The required data are  available for a sample of 103 transiting planets listed in the `The Extrasolar Planets Encyclopaedia' (Schneider 2011). The time to spiral in from a distance $a_{\rm 0}$ is given by (from eq.\ 4 in Debes \& Jackson)
\begin{equation}
t=a_0^{13/2}{4\over 117}({M_*\over G})^{1/2}{\qpe\over M_{\rm p}R_{\rm *}^5}.\label{tspiral}
\end{equation}
We define the  excess $\delta$ of the planet's radius $R_{\rm p}$ relative to its equilibrium radius  $R_0(M_{\rm p})$ as
\begin{equation} \delta=R_{\rm p}/R_0(M_{\rm p})-1.\end{equation} 
 For $R_0(M_{\rm p})$   we take  the mass-radius relation in Fortney et al. (2011). 
Fig.\ \ref{age1} shows the correlation between spiral-in life times $t$ thus calculated and the radius excess $\delta$ if a value   $\qpe = 10^6$ is assumed.  
The correlation between these two quantities does not depend on the value of \qp if the dissipation parameter is a constant. {\tt don't understand. what do you mean here?}

The spiral-in time could be an indicator of the planet's age by the statistical argument that a population of observed objects that are about to disappear in the near future is probably not very old. 
In Fig.\ \ref{histoage1} we show the age distribution obtained with a constant value of   $\qpe = 10^6$. This distribution would suggest that about half of the hot Jupiters have formed within the last 1 Gyr of 
the age of the Milky Way. 
However, the usefulness of spiral-in time as an age indicator is limited by its sensitivity to the value of   $\qpe$, which is highly uncertain. 
On the other hand,  the correlation shown in Fig.\ \ref{age1} could come from other effects, such as the trend that planets with stronger irradiation 
tend to have larger radii as discussed by Laughlin, Crismani \& Adams (2011), but this would not eliminate the conclusion that inflated planets would be young if {  $\qpe$ has values of about $10^6$ or lower.

\begin{figure}[t]
\hfil\includegraphics[width=0.8 \linewidth]{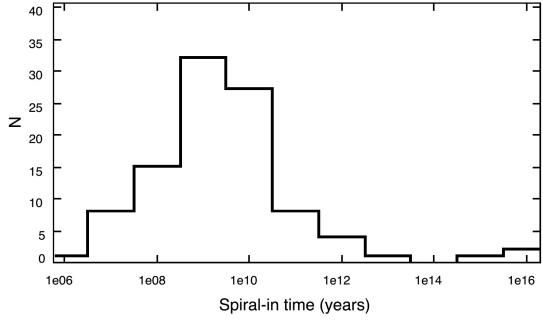}\hfil
\caption{Distribution of spiral-in times obtained from the results presented in Fig.\ \ref{age1}.}
\label{histoage1}
\end{figure}

Spiral-in times are  compared with standard cooling curves of irradiated planets in Fig.\ \ref{age2}. 
Since spiral-in time measures age only in an average sense, we have binned the data from Fig.\ \ref{age1}, 
sorting the sample by the value of $\delta$ in groups of 4 and averaging their values of $\delta$. 
The ages of the bins are calculated as geometric means. 
Curves show the predicted radius excess as a function of age from the irradiated models in Fig.\ 6 of Baraffe et al. (2003). 
The overall agreement between the time scales derived from the tidal evolution and from the cooling tracks is poor for the adopted   value of  $\qpe = 10^6$. 
A good agreement would require a value of   $\qpe = 3 \times 10^4$. This value is much lower than the single-value time-averaged estimates from ab-initio calculations by Ogilvie \& Lin (2007). 
More details on the dependence of tidal evolution with \qp can be found in Carone \& Patzold (2007).

\begin{figure}[t]
\hfil\includegraphics[width=0.8 \linewidth]{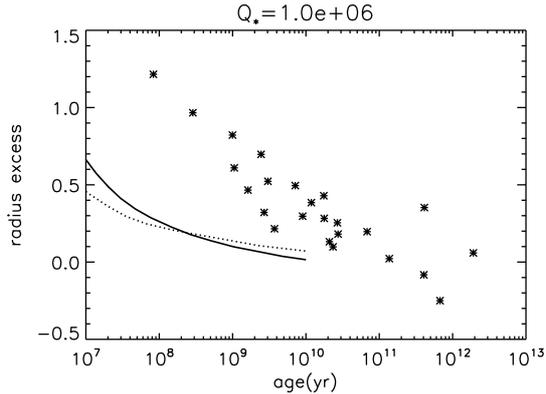}\hfil
\caption{Data of Fig.\ \ref{age1} binned and compared with planet cooling tracks from Baraffe et al. (2003). Solid: track for $10 M_{\rm J}$, dotted $1 M_{\rm J}$.}
\label{age2}
\end{figure}

\subsection{Host star lithium and rotation}

If the most inflated hot Jupiters are indeed the youngest planets in the sample, there could be correlations between radius excess and other age indicators such 
as the lithium abundance or rotation period of the host stars.  The possibility that hot Jupiters could be young has been discussed by Gandolfi et al. (2010).
 They find that the size of {/bf \object{CoRoT-11b}} fits standard cooling models if its age is only 12 Myr. They reject this possibility because CoRoT-11 has depleted its lithium. 
 The depletion of lithium is expected, however, if the star has been formed in a binary merger because simulations of stellar collisions indicate that there is 
 a substantial amount of chemical mixing (Trac, Sills \& Penn 2007). 
 The lithium of each of the stars in the contact binary would have been depleted during the pre-main sequence and main sequence evolution of the stars prior to the merging event  (cf.\  Martin 1997).
 Observations of cataclysmic binaries have indicated that lithium depletion is efficient during mass transfer evolution (Martin et al. 1995). 
 Thus, the binary merger scenario leads to the conclusion that the host stars of hot Jupiters should have enhanced lithium depletion, and this is consistent with 
 observational data (Israelian et al. 2004; Takeda et al. 2007). 
 
After the merger event, the resulting single star  would be a fast, magnetically active rotator surrounded by a disk. If the most inflated hot Jupiters are young, in the age range from 10 to 100 Myr, it could be expected that their host stars would be fast rotators because they still did not have enough time to slow down due to 
magnetic braking. Some examples of fast rotation have been noted in the literature, such as CoRoT-11 (Gandolfi et al. 2010) and 
HD 15082 (Collier Cameron et al. 2010). To check whether there is a correlation between planet radius excess and host star rotation we have plotted the data in Fig.\ \ref{rot}. No clear correlation is seen, but we note that all the planets with radius excess below about 20\% have host stars with rotation periods longer than five days. 

Though the hosts rotate more rapidly than normal stars of the same inferred age (Pont  2009, Hartman 2010), the difference is not as large as might be expected for post-merger ages of 10 -- 100 My. We tentatively suggest that the difference is due to the compact nature of a post-merger disk. Its specific angular momentum is much lower than in the large protostellar disks of T Tauris, and with a mass on the order of a tenth of a solar mass, it might be more efficient at braking the host star rotation through interaction with its magnetosphere.

\begin{figure}[t]
\hfil\includegraphics[width=0.9 \linewidth]{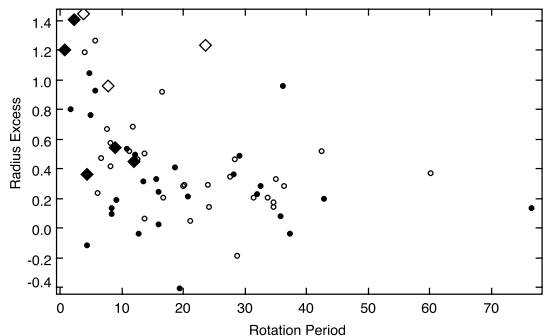}\hfil
\caption{ Radius excess $\delta$ vs.\ rotational  period (days)  for the host stars of 103 planets for which mass and radius of star and planet are given in the `Extrasolar Planets Encyclo\-paedia'.  
 Open diamonds: stars earlier than F5 with
planets with null orbital eccentricity; filled diamonds: stars earlier than F5 with significantly eccentric planet orbits; open circles:  stars later than F5 with planets with zero orbital eccentricity; filled circles stars later than F5 that host planets with significant eccentricities.}
\label{rot}
\end{figure}

\subsection{Blue stragglers}

The expected eventual merging of a W~UMa star would produce a single star that would look younger than a star of the same age as the original binary system. Such mergers are the default origin of the blue 
stragglers (BSS) in globular clusters (main sequence stars above the turn-off point), as has been proposed 
already by McCrea (1964) and studied by Mateo et al. (1990) and St\k{e}pie\'n (1995, 2009). Some BSS 
are found to be fast rotators, but the distribution of rotation velocities peaks at a modest $V_{\rm rot} \sin i=7$ km/s in the 47 Tuc globular cluster (Ferraro et al. 2006). In Fig.\ \ref{BSSrot} we compare the 
distribution of $v\sin i$ values measured for BSS in 47 Tuc with that of the transiting hot Jupiters for the
overlapping 0 mass range between 1.0 and 1.3 solar masses. A multiplicative correction factor was 
applied to the BSS data to account for their unknown inclination angle following the prescription of 
Chandrasekar \& Munch (1950). 
A two-sample Kolmogorov-Smirnov test indicates that the probability that these two populations are identical is 88.3\%. The statistical similarity is striking, although we note that there seems to be trend for higher $v\sin i$ values among the BSS in 47 Tuc. 
Thus, in line with the conclusions from the previous section, we identify the slow rotation of some hot Jupiter host stars as the main pitfall of the binary merger hypothesis. 

 \begin{figure}[t]
\hfil\includegraphics[width=0.9 \linewidth]{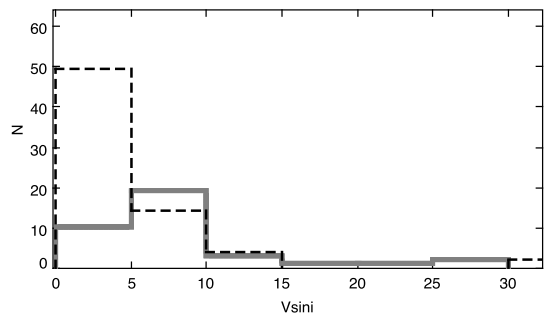}\hfil
\caption{Comparison of V sin$\i$ distributions  for host star of transiting hot Jupiters (dashed line) and BSS (solid line) in the 47 Tuc globular cluster for the stellar mass range between 
1.0 and 1.3 solar mass. A correction factor has been applied to account for the unknown inclination angle of the BSS.}
\label{BSSrot}
\end{figure}

In stellar mergers, accretion of chemically fractionated material may be enhanced over normal stars as shown by Desidera et al. (2007) in the case of HD 113984 AB, a wide binary  system where the primary 
is a BSS with an iron content 0.25 dex lower than the secondary star. A subpopulation of C and O 
depleted BSS has been identified in the 47 Tuc cluster by Ferraro et al. (2006), indicating again that 
stellar mergers can sometimes produce chemical abundance anomalies. 
 
\section{Final remarks}

We have presented the working hypothesis that hot Jupiter planets could form in binary mergers.  
The final merger of W~UMa stars is a possibility, but the direct merger of lower mass binaries that have not gone through an extended W~UMa-type contact phase 
is likely to contribute as well. We estimate that the frequency of these binaries in the Milky Way is consistent with what would be needed to account for 
the observed frequency of hot Jupiters. 

To test our hypothesis we looked for a correlation between the radius anomaly and the life time of the planet before spiral-in by tidal interaction with the host star. Using a  value   $\qpe = 10^6$ for the stellar 
dissipation constant  we found that about half of the hot Jupiters could have ages of about 1 Gyr or
younger. However, the usefulness of this age indicator is limited by its strong dependence on the value 
of   $\qpe$, which is highly uncertain.

A merger scenario agrees with the peculiar distribution of orbital distances of transiting planets. The concentration near 0.05 AU, with a rapid decline at larger distances, indicates an origin close to the host star. Merger events are likely to produce planets even closer in as well, but these would have disappeared rapidly from the population owing to the steep dependence of spiral-in time on distance. 

To  bring their radii in agreement with standard cooling curves (Baraffe et al. 2003), the most inflated planets would need to be younger  than about 0.1 Gyr. A correlation between degree of inflation and the 
rotation rate of the host star would be expected from the merger scenario. It is found to be only weakly 
present in the observations, overshadowed by large scatter. In agreement with the scenario, however, the 
average rotation rate of the hosts of inflated planets is greater than in normal stars of the same (inferred) 
age, as noted previously (Pont 2009, Hartman 2010).
  
The inclined and retrograde orbits identified through the Rossiter-McLaughlin effect (cf.\  Triaud et al. 2010) 
require  explanation in the merger interpretation, since the angular momentum of a disk formed in a merger would have the same direction as that of the star. We suggest here that planet formation in the massive, compact disk resulting from merger on a dynamical time scale, would have produced a number of planets in closely packed orbits. Gravitational interaction in such compact systems of planets causes their orbits to be secularly unstable. This would scatter them into stochastic orbits (e.g.\ Chatterjee et al. 2008), including retrograde ones. (The total angular momentum vector and a corresponding preponderance of prograde orbits would be conserved). The process is also consistent with the high frequency of excentric orbits. 
 
Main sequence GK stars with super-solar metallicity have a markedly increased frequency of hot Jupiters 
(Gonzalez et al. 2001; Santos et al. 2004; Fischer  \& Valenti 2005) 
 and a similar trend is seen in M dwarfs (Johnson \& Apps 2009).
 Another intriguing observation is the presence of a massive planet around a very metal-poor star 
 (Setiawan et al. 2011).These strange observational results could be explained in the context of stellar 
 mergers, which may lead to unexpected abundance patterns in the resulting host stars and 
will form planets even around the most metal-poor stars in the Universe.

The frequency of stellar companions around contact binaries has been found to be 31\% $\pm$6 \% using 
adaptive optics assisted imaging (Ruckinski, Pribulla \& van Kerkwijk 2007). 
in the separation range of 3 to 100 AU. The survival of these companions will depend on their mass, their 
orbital separation, and the amount of mass loss from the system during 
the merger event. The evolution of these triple systems needs to be modeled to be able to  compare with 
the binary properties of hot Jupiter host stars so that a useful constrain can be derived.  

Our main conclusion is that the hypothesis that hot Jupiters may result from binary mergers has some advantages over other competing explanations of the radius anomaly, and it may also 
account for other odd observational trends. However, it has a clear issue with the slow rotation of host stars of inflated planets. 
We note that it would be worthwhile to study the detailed mechanisms of angular momentum loss and disk formation in binary mergers. 

As observational tests of the binary merger scenario for the formation of hot Jupiters we propose the following: a)  the frequency of these planets among BSS should be significantly higher than among 
T Tauri stars and b) the  age of some exoplanet hosts inferred from nucleocosmochronology could be older than the age estimated from standard evolutionary models for 
single stars.

\section*{Acknowledgments}
\bhl We thank the referee for detailed and constructive criticism.\ehl
This work has been supported by the Spanish Plan Nacional de Astronomia y Astrofisica under grants AYA2007-2357 and AYA 2010-21308-C03-02, by RoPACS, a Marie Curie 
Initial Training Network funded by the European Commission's Seventh Framework Program, and by the CONSOLIDER-INGENIO GTC project.  
We thank Michele Montgomery for detailed comments on an earlier version of the text.
Part of this work was carried out while EM was a visiting research professor at the Geological Sciences Department of the University of Florida in Gainesville.

\clearpage

\end{document}